\documentclass[useAMS,usenatbib,usegraphicx]{mn2e}
\usepackage{xcolor}
\onecolumn{}

\title[WASP 1021-28]{Pulsations and Pre-He White Dwarf in the Post-mass Transfer Eclipsing System WASP 1021-28}

\author[J. W. Lee et al.]
       {Jae Woo Lee$^{1}$\thanks{E-mail: jwlee@kasi.re.kr}, Min-Ji Jeong$^{1,2}$ and Kyeongsoo Hong$^{1}$ \\
        $^1$Korea Astronomy and Space Science Institute, Daejeon 34055, Republic of Korea \\
        $^2$Department of Astronomy and Space Science, Chungbuk National University, Cheongju 28644, Republic of Korea}

\begin{document}

\date{Accepted 2025 ---------. Received 2025 ---------; in original form 2025 }

\pagerange{\pageref{firstpage}--\pageref{lastpage}} \pubyear{2025}

\maketitle

\label{firstpage}

\begin{abstract}
We present results from VLT/UVES spectra and TESS photometric observations of the pulsating EL CVn binary WASP 1021-28, 
containing a He-core white dwarf precursor (pre-He WD). Double-lined radial velocities were measured with the atmospheric 
parameters of $T_{\rm eff,A}$ = 7411$\pm40$ K, [M/H] = 0.34$\pm$0.05 dex, and $v_{\rm A}$$\sin i$ = 86.6$\pm$4.0 km s$^{-1}$ 
for the more massive primary. Combining these measurements and TESS data from four sectors allowed the direct calculation 
of accurate values for the absolute parameters of each component and the distance to the system. The third-light source 
of $l_3$ = 0.029 may be the outer tertiary object previously discovered by SPHERE/IRDIS observations. 
WASP 1021-28 A is located near the blue edge of the $\gamma$ Dor instability strip, and the less massive companion is concurrent 
with the He-core WD model for metallicity $Z$ = 0.02 and mass $M$ = 0.191 $M_\odot$. The $Z$ value and the Galactic kinematics 
demonstrate that the program target belongs to the thin-disk population. We iteratively prewhitened the entire TESS residuals and 
extracted four and nine significant signals in two ranges of 1.12$-$2.25 day$^{-1}$ and 111.25$-$139.24 day$^{-1}$, respectively. 
A signal of $f_2$ = 1.31865 day$^{-1}$ in the low-frequency region can be attributed to the $\gamma$ Dor pulsation of WASP 1021-28 A, 
and the high frequencies may be extremely low-mass pre-He WD oscillations. The results presented here provide valuable information 
on the evolution of short-period EL CVn stars proposed as inner binaries of hierarchical triple systems and the multiperiodic pulsations. 
\end{abstract}


\section{INTRODUCTION}

This study is a continuation of a series of papers that measure the fundamental parameters and pulsation frequencies for each 
component of the EL CVn-type binaries, thereby understanding their physical properties and evolutionary states (see Lee et al. 2024). 
The eclipsing binaries (EBs) are post-mass transfer systems, comprised of a main-sequence (MS) star of spectral type A or F and 
an extremely low-mass white dwarf precursor (pre-ELM WD) with a mass below 0.3 M$_\odot$. The pre-ELM WDs are known to be 
the result of non-conservative mass transfer evolution during the red-giant branch (RGB) phase of a binary system, rather than 
a single star, given the age of the Universe. They are the remnants of helium (He) cores that evolved from the more massive 
component in the initial MS binaries, and are called proto-/pre-He WD. Comprehensive reviews of the EL CVn stars appear in 
Maxted et al. (2014a), van Roestel et al. (2018), and Lee et al. (2020). To further develop this subject, we explore another such 
EL CVn binary, WASP 1021-28 (1SWASP J102122.83-284139.6; TIC 192990023, TYC 6631-538-1, 2MASS J10212283-2841395, Gaia DR3 5462478705727653248; 
$V$ = $+$11.12, $(B-V)$ = $+$0.39; $H$ = $+$10.43, $(J-H)$ = $+$0.07).  

The target star was announced as an EL CVn-type EB by Maxted et al. (2014a) based on archival data from the WASP survey. Using 
the light curve fitting, they reported that WASP 1021-28 is a well-detached stellar system with the following binary parameters: 
inclination angle $i$ = 82.1$\pm$0.6 deg, photometric mass ratio $q_{\rm ph}$ = 0.143$\pm$0.027, relative radii $r_{\rm A}$ = 0.3707$\pm$0.0034 
and $r_{\rm B}$ = 0.0767$\pm$0.0010, and luminosity ratio $L_{\rm B}/L_{\rm A}$ = 0.1026$\pm$0.0011. Maxted et al. (2014a) also 
compared model and observed flux distributions that lead to effective temperatures of $T_{\rm eff,A}$ = 7300$\pm$300 K and 
$T_{\rm eff,B}$ = 9800$\pm$500 K. The subscript A denotes the larger and brighter primary star eclipsed in the secondary minimum, 
and the hotter companion is indicated by subscript B. 
On the other hand, Lagos et al. (2020) discovered an additional object, from VLT/SPHERE observations, that is $\sim$1.0 arcsec 
away from the EB and fainter by $\Delta H$ = 3.595$\pm$0.008 mag, using the IRDIS DBI mode in the $H$ filter and the 185-mas 
diameter coronagraph N$\_$ALC$\_$YJH$\_$S (Beuzit et al. 2008; Dohlen et al. 2008; Vigan et al. 2010; Guerri et al. 2011). 
They estimated the mass and temperature of the tertiary object to be about 0.6 M$_\odot$ and 3700$-$4000 K, respectively, 
corresponding to a spectral type of approximately K9 V (Pecaut \& Mamajek 2013). The probability of being aligned by chance 
with a background source is only 0.04\%.

While writing this paper, \c Cakirli et al. (2024) reported an analysis of the same VLT/UVES spectra and TESS light curve that 
we used. We believe that our results will help to better understand the absolute properties of WASP 1021-28 and 
the multiperiodic oscillations excited in each component. In this paper, we compare their results with ours and present in detail 
the pulsation features and evolutionary state of the double-line EL CVn EB.

\section{TESS PHOTOMETRY AND ECLIPSE TIMES}

WASP 1021-28 was observed as part of the TESS photometric survey (Ricker et al. 2015) to search for small transiting exoplanets. 
The satellite observations were conducted in Sectors 9 (S9), 36 (S36), 62 (S62), and 63 (S63) between 2019 March and 2023 April. 
The S9 data were acquired in 2-min sampling mode only, while the other ones were collected at both 2-min and 
20-s cadences. We downloaded the TESS observations from the MAST archive\footnote{https://archive.stsci.edu/}, which are 
available at http://dx.doi.org/10.17909/0d6n-4767. The simple aperture photometry data (\texttt{SAP$_{-}$FLUX}) were 
processed and used in this study employing the same approach as previously applied (Lee et al. 2017, 2024). To convert the flux 
to magnitude units, we took the TESS magnitude at maximum light to be $T_{\rm p}$ = $+$10.871 (Paegert et al. 2022). 
The resulting time-series data are presented in Figure 1. The total time span of these observations is about 1490 days, with 
observation gaps of 714 and 684 days between S9 and S36 and between S36 and S62, respectively. The average of the CROWDSAP factors 
reported in the four sectors is 0.9915$\pm$0.0030, which suggests that the TESS measurements of WASP 1021-28 are hardly influenced 
by surrounding sources. 

To measure the eclipse mid-times of WASP 1021-28, we applied the Kwee \& van Woerden (1956) method to each eclipse curve in 
the TESS data. In total, 91 primary and 90 secondary minimum epochs were extracted from the 2-min (S9) and 20-s (S36, S62-63) 
cadences, which are summarized in Table 1. A linear least-squares fit to the primary minima yielded the following orbital ephemeris 
for the TESS data of WASP 1021-28: 
\begin{equation}
\mbox{Min I} = \mbox{BJD}~ 2,458,545.09331(\pm0.00012) + 0.900898357(\pm0.000000087)E, 
\end{equation}
where the numbers in parentheses are the 1$\sigma$-error values. The TESS orbital period ($P_{\rm orb}$), corresponding to 
the frequency $f_{\rm orb}$ = 1.1100031 $\pm$ 0.0000001 day$^{-1}$, is slightly longer than the value (0.9008980$\pm$0.0000002 days) 
derived from the WASP archive data (Maxted et al. 2014a).

\section{VLT/UVES SPECTROSCOPY AND DATA ANALYSIS}

High-resolution spectroscopy of WASP 1021-28 was performed between 2014 November 22 and 2015 February 17 in the context of 
ESO under programme 094.D-0027(A), which was devoted to observing EL CVn-type binaries containing pre-He WDs. The spectroscopic 
observations were made with the UVES Spectrograph (Dekker et al. 2000) attached to the VLT 8.2-m telescope at Paranal in Chile. 
The blue (wavelength range 3282--4563 \AA) and red (wavelength range 5655--9464 \AA) arms with slit widths of 0.8 arcsec were 
adopted, providing dispersions of 49,620 and 51,690, respectively. The observation instrument setup and data reduction for 
the target star are identical to those for WASP 0346-21 (Lee et al. 2024). A total of seven reduced spectra were acquired at 
the ESO Science Archive Facility\footnote{https://archive.eso.org/scienceportal/home} and utilized in our study. 
They consist of three observed spectra, each with an integration time of 120 s, most of which have a signal-to-noise ratio (SNR) 
of about 30. We normalized the archival data by applying a spline3 function. 

The UVES spectral analysis was carried out in the same manner as in the study by Lee et al. (2024). For the RV measurements 
of WASP 1021-28, we applied the broadening function (BF) formalism (Rucinski 2002) built in the RaveSpan software 
(Pilecki et al. 2017) to the normalized echelle spectra. This method is particularly useful for multiple or binary systems 
where the rotational broadening between the component stars differs significantly, and can provide better resolution profiles 
than the cross-correlation function (Rucinski 2002, 2004). In this analysis, we adopted a template spectrum from the model grids 
of LTE synthetic spectra by Coelho et al. (2005). 

The low SNR spectral lines of WASP 1021-28 B can be challenging to detect because of its low luminosity. We scrutinized 
the trailed spectra to identify the spectral regions where the absorption lines of both components are observable 
(e.g., Lee et al. 2023). Our examination uncovered two specific regions of 4440--4520 \AA and 6320--6405 \AA, which clearly 
show the absorption lines of both components moving along the binary's orbital motion. We also applied the BF method to regions 
excluding Balmer lines with wide wings and where telluric lines are sparse, and found binary features in the region 4160--4280 \AA, 
in addition to the two ranges previously revealed. The observation times of the blue- and red-arm data are less than 1 second 
apart, and the preliminary measurement errors of RVs are larger than the velocity change caused by the time difference. 
Therefore, to extract the RVs of the EB components, three spectral ranges of 4160--4280 \AA, 4440--4520 \AA, and 6320--6405 \AA 
were analyzed simultaneously using rotation and Gaussian profile functions separately. 

Figure 2 shows the sample BF profiles at quadrature phases, where the rotation functions in the left panels describe the BFs 
much better than the Gaussian functions in the right panels. Consequently, we chose the RV measurements obtained from the rotation 
profile fits as the final values presented in Table 2 and Figure 3. \c Cakirli et al. (2024) also adopted the BF method of 
the RaveSpan to obtain the UVES RVs, but applied a Gaussian function to each spectral signal instead of a rotation function. 
Our and their RVs differ on average by $\Delta V_{\rm A}$ = $2.07\pm0.55$ km s$^{-1}$ and $\Delta V_{\rm B}$ = $-0.25\pm3.98$ km s$^{-1}$ 
for the A and B components, respectively, and these errors are expressed as 1$\sigma$ values from the standard deviations. 

The atmospheric properties of WASP 1021-28 A were derived using the grid search software package GSSP\footnote{https://fys.kuleuven.be/ster/meetings/binary-2015/gssp-software-package} 
(Tkachenko 2015), which compares the observed target spectra with grids of synthetic model spectra and minimizes the $\chi^2$ 
statistic between them. For this, we utilized the UVES spectrum of the orbital phase 0.983 (BJD 2,456,983.82076) obtained during 
the primary eclipse. The synthetic model grids were generated over the ranges of effective temperature $T_{\rm eff}$ = 
6800$-$8000 K, metallicity [M/H] = $-$0.5$-$+0.5 dex, and projected rotational velocity $v_{\rm A}$sin$i$ = 80$-$120 km s$^{-1}$. 
In these grids, the surface gravity, and micro- and macro-turbulence velocities were set to $\log$ $g_{\rm A}$ (cgs) = 4.16 
(see Section 4), and $v_{\rm micro}$ = 2.0 km s$^{-1}$ and $v_{\rm macro}$ = 19.9 km s$^{-1}$ from the iSpec code (Blanco-Cuaresma et al. 2014). 
The GSSP runs focused on the wavelength range 3700--4200 \AA, which includes absorption lines that are very sensitive to 
atmospheric parameter estimations. The grid search results yielded the optimal values for three atmospheric parameters: 
$T_{\rm eff,A}$ = 7411$\pm40$ K, [M/H] = 0.34$\pm$0.05 dex, and $v_{\rm A}$$\sin i$ = 86.6$\pm$4.0 km s$^{-1}$. Figure 4 displays 
our best-fit synthetic model superimposed on the observed spectrum at the conjunction phase.

Our $T_{\rm eff,A}$ value is in close agreement with the effective temperatures of 7300$\pm$300 K, 7378$\pm$140 K, and 7294$\pm$41 K 
from Maxted et al. (2014a), the TESS Input Catalogue (Paegert et al. 2022), and the Gaia DR3 archive (Gaia Collaboration 2022), 
respectively. In contrast, \c Cakirli et al. (2024) reported atmospheric parameters of 7650$\pm50$ K, 0.20$\pm$0.03 dex, 
and 82$\pm$2 km s$^{-1}$ in the same order as before using the iSpec code. While their metallicity and rotational rate match 
our measurements within the range of 3$\sigma$ errors, the temperature deviates significantly from all other measurements, 
including ours.

\section{BINARY MODELING} 

As stated in the Introduction, EL CVn stars are very promising binaries for investigating the absolute properties and 
evolutionary history of pre-ELM WDs. These studies require reliable measurements of the mass, radius, temperature, and 
luminosity of each binary component, which are achieved using precise light and spectral data. In particular, the mass ratio 
($q$), a key parameter in binary star modeling, can be accurately measured directly from the double-lined RVs. 
The TESS observations in Figure 1 are typical EL CVn light curves showing both a total eclipse at the primary minimum and 
an out-of-eclipse ellipsoidal variation, with no notable differences between the sectors utilized. For a unique solution for 
WASP 1021-28, we adopted the EB modeling code (hereafter W-D code) that was developed by Wilson \& Devinney (1971) and steadily 
improved over the past 50 years (see van Hamme \& Wilson 2007; Kallrath 2022). The full 2-min TESS light curves were analyzed 
with the UVES RV measurements in the same way as for WASP 0346-21 (Lee et al. 2024). 
The mean time difference between the primary and secondary minimum epochs in the TESS data is 0.45048$\pm$0.00049 days. 
This value corresponds to half the orbital period, suggesting that the EB target is likely in a circular orbit ($e$ = 0.0). 

From our spectral analysis, we set the effective temperature of WASP 1021-28 A to $T_{\rm eff,A}$ = 7411$\pm40$ K and 
its rotation parameter to $F_{\rm A}$ = 0.890$\pm$0.042, which is the ratio of the observed $v_{\rm A}\sin$$i \approx v_{\rm A}$ 
and the synchronous rotation $v_{\rm A,sync}$. The bolometric albedos and gravity-darkening coefficients of the component stars 
were given as $A_{\rm A,B}$ = 1.0 and $g_{\rm A,B}$ = 1.0, respectively, as expected for the temperatures. 
The logarithmic limb-darkening law was chosen, and its parameters $x_{\rm A,B}$ and $y_{\rm A,B}$ were fixed from 
the van Hamme (1993) tables provided in the W-D code. WASP 1021-28 B was assumed to be in a rotation state synchronized with 
the system's orbital motion. The spectroscopic orbital elements of the mass ratio, semi-major axis, and systemic velocity were 
initialized to $q$ = 0.116, $a$ = 4.77 R$_\odot$, and $\gamma$ = 6.6 km s$^{-1}$ by a combination of our sinusoidal fits to 
RV curves and the photometric solution of Maxted et al. (2014a). Other adjustable parameters in this modeling are 
the orbital epoch ($T_0$) and period ($P_{\rm orb}$), the inclination ($i$), the surface potentials ($\Omega_{\rm A,B}$) of 
both stars, the secondary's temperature ($T_{\rm eff,B}$), the primary's luminosity ($l_{\rm A}$), and the third light ($l_3$). 

Most short-period EL CVn EBs are proposed to be inner close pairs of hierarchical triples that host circumbinary companions 
(Lagos et al. 2020; Lee et al. 2024). The TESS light and UVES RV measurements of WASP 1021-28 were not simultaneous and 
the time interval between them is more than 4 yr. Therefore, unlike the simultaneous analysis of \c Cakirli et al. (2024), 
we modeled the two types of observations separately and repeatedly. First, we solved the 2-min sampling TESS data using 
the initial spectroscopic elements. Then, the double-lined RV curves were solved applying the photometric parameters computed in 
the previous step. We re-performed the light curve analysis with the updated RV parameters. This procedure was detailed in 
Lee et al. (2024), and the final results are presented in Table 3 along with those of \c Cakirli et al. (2024). The model fits 
to the light and RV curves are plotted as the solid lines in Figures 1 and 3, indicating that the binary star model represents 
both datasets well. Our parameter errors were obtained via the error estimation method of Southworth \& Bowman (2022), which 
utilizes the variation of each free parameter from various model fits using different inputs and alternative approaches.

Looking at Table 3, the modeling parameters that show the largest difference between our and \c Cakirli et al. (2024) solutions 
are the luminosities ($l_{\rm A}$ and $l_3$) contributed by each component to the total brightness of the potential triple star 
in the TESS band. Our third light ($l_3$ = 0.029) is about 4 times smaller than that ($l_3$ = 0.120) of \c Cakirli et al. (2024), 
resulting in an orbital inclination that is about 2 deg smaller. The light contribution of 2.9 \% is in satisfactory agreement 
with that ($\sim$3.5 \%) obtained from the magnitude difference of $\Delta H$ = 3.595$\pm$0.008 mag between WASP 1021-28 AB and 
the tertiary object detected by Lagos et al. (2020). The $l_3$ source could then be a distant outer tertiary observed in 
the SPHERE/IRDIS $H$ band images, gravitationally bound to the eclipsing pair of WASP 1021-28. On the other hand, our errors in 
the light curve parameters are generally higher than those of \c Cakirli et al. (2024), while the errors in the velocity parameters 
are about one order of magnitude smaller, and so is the mass ratio ($q$ = $K_{\rm A}$/$K_{\rm B}$). Regarding the spectroscopic 
orbits, the main difference between us and them is the method for measuring the RVs. The BF profiles for the UVES spectra were better 
described by a rotational function than by a Gaussian function, so we prefer our RV measures. 

The absolute parameters of WASP 1021-28 were derived using our light and velocity solution, and are summarized at the bottom of 
Table 3. In this calculation, the effective temperature and bolometric magnitude of the Sun were given as $T_{\rm eff}$$_\odot$ 
= 5780 K and $M_{\rm bol}$$_\odot$ = +4.73, and the bolometric correction (BC) of each component was obtained through 
temperature calibration (Flower 1996, Torres 2010). We also determined the distance of WASP 1021-28 as 558$\pm$9 pc by applying 
the apparent maximum light and color excess of $V$ = 11.115$\pm$0.011 and $E$($B-V$) = 0.039$\pm$0.009 (Paegert et al. 2022) to 
the well-known equation of $V-A_{\rm V}-M_{\rm V,EB}=5\log{d}-5$, where the interstellar extinction $A_{\rm V} \simeq 3.1E(B-V)$ 
and the EB's total absolute magnitude $M_{\rm V,EB}$ = 2.291$\pm$0.034. This distance represents a parallax of 1.792$\pm$0.029 mas, 
which matches well with the $Gaia$ measurement of 1.807$\pm$0.020 mas (Gaia Collaboration 2022), corresponding to a distance of 
553$\pm$6 pc.

\section{PULSATIONAL CHARACTERISTICS}

The fundamental parameters of WASP 1021-28 indicate that the more massive A component is a normal MS of spectral type A9, while 
the hotter B companion is a He-core WD precursor. Thus, the two components are likely to be a $\delta$ Sct/$\gamma$ Dor variable 
and a pulsating WD in the same order (Maxted et al. 2013, 2014b; Wang et al. 2020; Hong et al. 2021; Lee et al. 2023, 2024). 
\c Cakirli et al. (2024) analyzed separately three TESS datasets (S9, S36, S62$-$32), separated by observational gaps. Using 
the SNR$\geq$4 criterion, they found 17 frequencies in S9, 15 in S36, and 14 in S62$-$63. However, only three frequencies of 
24.4 day$^{-1}$, 112.4 day$^{-1}$, and 126.7 day$^{-1}$ were detected together in all three datasets, the first of which is 
22 times the orbital frequency (22$f_{\rm orb}$). Our frequency search for WASP 1021-28 was conducted using the software PERIOD04 
(Lenz \& Breger 2005) with successive prewhitening (Lee et al. 2014).

For a useful Fourier analysis, we used the TESS residual lights cleaned with the W-D model curve. Analysis of the model-subtracted 
data revealed several peak signals that were multiples of the orbital frequency $Nf_{\rm orb}$, where $N$ is an integer. Since 
WASP 1021-28 favors a circular orbit in both the TESS observations and binary modeling, it is difficult to attribute these 
harmonics to tidal excited pulsations, which would typically occur in binary systems with high eccentricities of $e>0.2$ (Guo 2021). 
The orbital harmonic oscillations may primarily result from systematic trends and orbital effects not being completely removed 
from the observed TESS data. Therefore, we first removed the orbital frequencies from the light curve residuals, and then conducted 
the prewhitening process on the immune data up to the Nyquist limit $f_{\rm Ny}$. 

Figure 5 shows the amplitude spectra for the entire 2-min sampling data of S9, S36, S62, and S63 from the PERIOD04. Here, 
the frequency signals are clearly visible in two main parts: the low-frequency region $< 10$ day$^{-1}$ and the high-frequency 
region 100$-$150 day$^{-1}$. In contrast, there was no noticeable signals in the frequency range over 150 day$^{-1}$. 
Our repeated prewhitening continued until no frequency had an SNR greater than $\sim$5 
(Baran \& Koen 2021; Bowman \& Michielsen 2021). As a consequence, we found 11 frequency peaks (4 low and 7 high) in 
the full-phase 2-min cadences. Frequencies between 13 and 45 day$^{-1}$ found in each dataset by \c Cakirli et al. (2024) were 
either removed as orbital harmonics or did not meet the SNR$\ga$5 criterion. 

In the 3rd and 4th panels of Figure 1, the scatter band of the 2-min cadence residuals is about half that of the 20-s ones, which 
means that the former data is more precise than the latter and can better detect relatively low frequencies such as $\delta$ Sct 
and $\gamma$ Dor pulsations. Nevertheless, the ultra-short cadences of 20 s would be more useful for uncovering the high-frequency 
signals from the pre-He WD. We introduced PERIOD04 for all 20-s residual lights except for the primary eclipse time data, 
where the B component is completely obscured by the MS star. In this analysis, our interest was focused on the oscillations of 
the pre-He WD companion with frequencies higher than 70 day$^{-1}$. The PERIOD04 periodograms for the 20-s cadences with 
prewhitening applied are illustrated in Figure 6. We found two additional frequencies (121.33540 day$^{-1}$ and 138.30988 day$^{-1}$) 
in the analysis of 20-s sampling data compared to the 2-min data analysis. Our final results are summarized in Table 4, ordered 
by frequency. 

When analyzing data from multiple sectors, the long-term observation gaps between them can introduce artifact 
frequencies in the low frequency region due to uncorrected trends in the data. In order to validate the low frequencies detected 
in the full 2-min cadence data of WASP 1021-28, we analyzed the residual lights for each sector separately. The amplitude spectra 
of each of the four sectors (S9, S36, S62, and S63) are presented in Figure 7, where the gray and red lines represent 
the periodograms for the W-D model residuals and the immune data with orbital frequencies removed from them, respectively. 
In this figure, the $f_2$ frequency is present in all sectors and its SNR is greater than 4 in three sectors except S63. 
In contrast, the other three low frequencies are absent in any sector and are thought to be aliasing sidelobes derived from 
the orbital frequencies. The result of this process indicates that the low frequency of $f_2$ = 1.31865 day$^{-1}$ is likely 
a $\gamma$ Dor oscillation of WASP 1021-28 A and the high frequencies are pre-ELM variables associated with its hot companion. 
However, the low-frequency signal is not dominant compared to its surrounding noise peaks, so our interpretation of $f_2$ is 
preliminary and requires further verification.

\section{DISCUSSION AND CONCLUSIONS}

We present the UVES spectra and TESS photometric data of WASP 1021-28 to characterize a post-mass transfer EB, consisting 
of an A-type MS primary and a pre-He WD, and the pulsations in each component. The double-lined RVs were derived from a rotation 
function fitted to each BF profile rather than a Gaussian function, and the atmospheric parameters for the A component were 
computed with the GSSP package from the observed spectrum at a conjunction phase of 0.983, as follows: $T_{\rm eff,A}$ = 7411$\pm40$ K, 
[M/H] = 0.34$\pm$0.05 dex, and $v_{\rm A}$$\sin i$ = 86.6$\pm$4.0 km s$^{-1}$. The spectroscopic measurements were modeled in 
combination with the full 2-min TESS data. The modeling result, summarized in Table 3, provided the best-fit solution for 
the light and RV curves, and thus the absolute dimensions for each component. The mass and radius of the primary component were 
derived to within 1 \% precision, and those of the less massive companion to about 1 \% and 2 \%, respectively. The observed 
$v_{\rm A}\sin$$i$ indicates that WASP 1021-28 A is currently in a sub-synchronous state. 

The absolute parameters of WASP 1021-28 were used to trace the evolutionary sequence and current state of the EL CVn binary in 
the Hertzsprung-Russell (H-R, upper) and $\log T_{\rm eff}-\log g$ (lower) diagrams of Figure 8. Here, the EB components are 
presented as cyan and pink star symbols. The oblique blue and red lines in the H-R diagram represent the empirical instability regions 
for the $\delta$ Sct (solid) and $\gamma$ Dor (dashed) variables, which indicates that WASP 1021-28 A is a candidate intermediate-mass 
pulsator located on the zero-age MS. In these diagrams, the black dotted, dashed, and solid curves display the evolutionary sequences 
of the ELM WDs with metallicities and masses ($Z$, $M/M_\odot$) of (0.001, 0.190), (0.01, 0.192), and (0.02, 0.191), respectively 
(Istrate et al. 2016). As can be seen, WASP 1021-28 B matches well with the WD model for $Z$ = 0.02 and $M$ = 0.191 $M_\odot$ and 
may be a precursor of a He-core WD evolving toward higher temperature at the midpoint of a constant luminosity stage. 

To assign the population membership of WASP 1021-28, we calculated the space motion of ($U, V, W$)\footnote{Positive toward 
the Galactic center, Galactic rotation, and North Galactic Pole.} = ($4.13\pm0.08$, $215.4\pm0.1$, $-10.8\pm0.2$) km s$^{-1}$ 
and the Galactic orbit of ($J_{\rm z}$, $e$)\footnote{Angular momentum in the $z$ direction and eccentricity} = 
($1726\pm19$ kpc km s$^{-1}$, $0.0609\pm0.0002$), following the same procedure and solar values as Lee et al. (2020). 
The kinematic data for WASP 1021-28 are located near the center of the thin-disk regions in the $U-V$ and $J_{\rm z}-e$ diagrams 
presented by Pauli et al. (2006). The population classification agrees well with the $Z$ = 0.02 value obtained by comparing 
our absolute properties with ELM WD models, indicating that WASP 1021-28 is a member of the thin-disk population.

Lagos et al. (2020) reported the presence of a new K-type star at 1.0 arcsec from WASP 1021-28 AB using the high-contrast 
SPHERE/IRDIS images in the $H$ band. However, we did not detect such an object in the UVES spectra, excluding the EB components. 
This non-detection is due to the fact that the spectroscopic observations were made with a slit width of 0.8 arcsec, which is 
narrower than the separation between the eclipsing pair and the tertiary object. In the binary modeling, we conducted 
the $l_3$ search and detected a third light of 2.9 \%, which is significantly different from that (12 \%) of \c Cakirli et al. (2024) 
but consistent with the light contribution from the potential tertiary detected by Lagos et al. (2020). As presented in Section 2, 
the CROWDSAP value of 0.9915$\pm$0.0030 for WASP 1021-28 indicates that most of the TESS flux measurements come from the target 
star, with the remaining approximately 1 \% being contaminated. Our $l_3$ source is then thought to be the outer tertiary of 
Lagos et al. (2020) around WASP 1021-28 AB. The circumbinary object would have removed most of the angular momentum from 
the early MS binary and contributed to its evolution into the current close pair with a short period. 
Moreover, the sub-synchronous rotation revealed in this paper may be affected by the perturbations of the circumbinary companion 
(Fuller \& Felce 2023). All of these indicate that WASP 1021-28 is a post-mass transfer EB, with a distant $K$-type tertiary and 
characterized by multiperiodic pulsations, making it a good benchmark target for studying the evolution and asteroseismology of 
EL CVn stars.

\section*{Acknowledgments}
This paper includes VLT/UVES spectra made with ESO under programme 094.D-0027(A) and photometric data collected by the TESS mission. 
We wish to thank David E. Mkrtichian for his careful reading and valuable comments on the frequency analysis. This research has made 
use of the Simbad database maintained at CDS, Strasbourg, France, and was supported by the KASI grant 2025-1-830-05. M.-J.J. was 
supported by the grant number RS-2024-00452238 from the National Research Foundation (NRF) of Korea.

\section*{DATA AVAILABILITY}
The data underlying this article will be shared on reasonable request to the first author.

\clearpage
\begin{figure}
\includegraphics[]{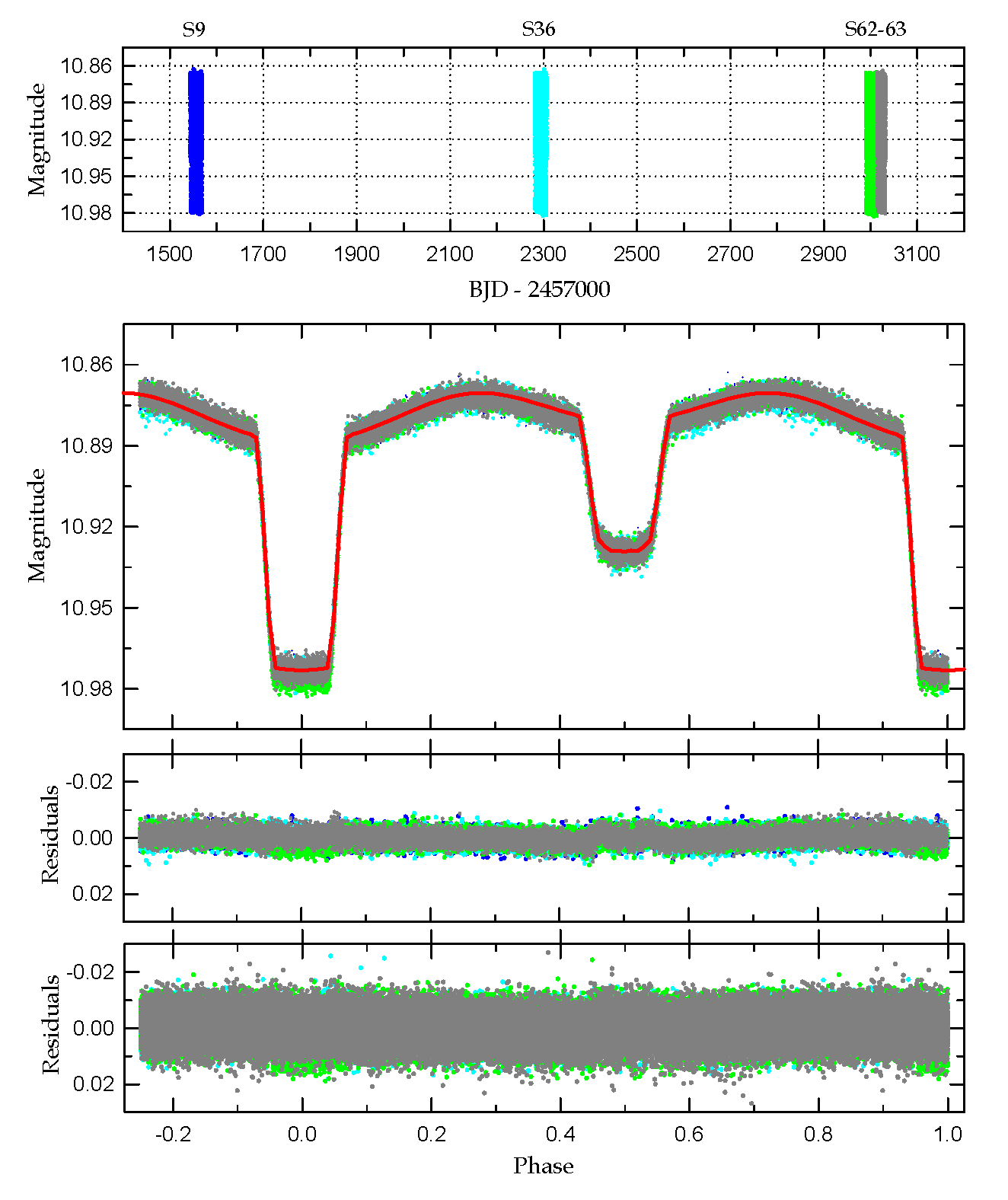}
\caption{TESS observations of WASP 1021-28 distributed in BJD (top panel) and orbital phase (second panel). The different colored 
circles for each sector are individual measures, and the red solid curve represents the synthetic model obtained from our W-D fit. 
The third and bottom panels show the corresponding residuals for 2-min and 20-s cadence data, respectively, from the model curve. } 
\label{Fig1}
\end{figure}

\begin{figure}
\includegraphics[]{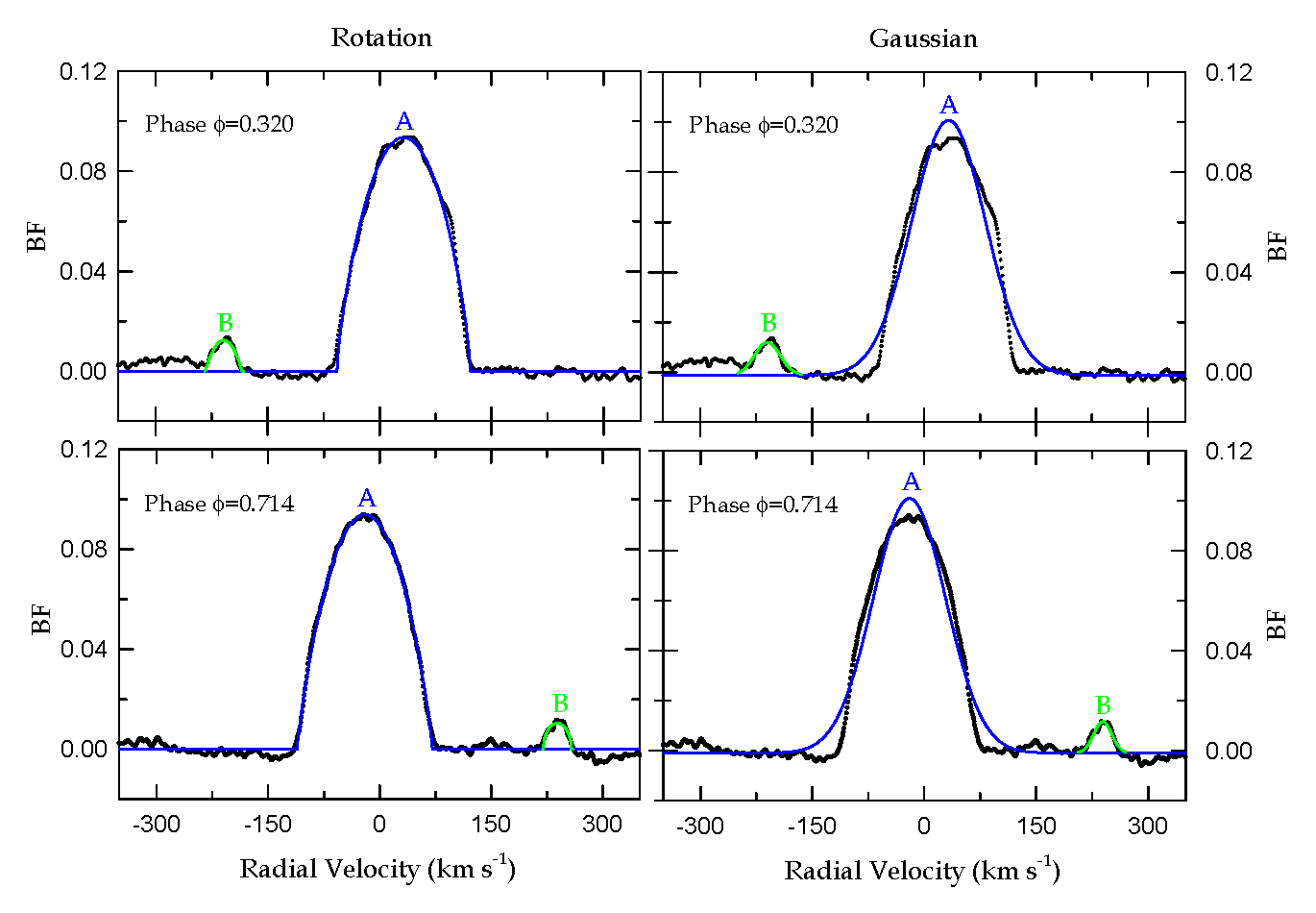}
\caption{Sample BF profiles for two orbital phases ($\phi$) obtained with the RaveSpan code. The black dots are the BFs with 
rotation (left panels) and Gaussian (right panels) functions applied. The blue and green lines represent the primary (A) and 
secondary (B) stars from the resulting profiles, respectively. }
\label{Fig2}
\end{figure}

\begin{figure}
\includegraphics[]{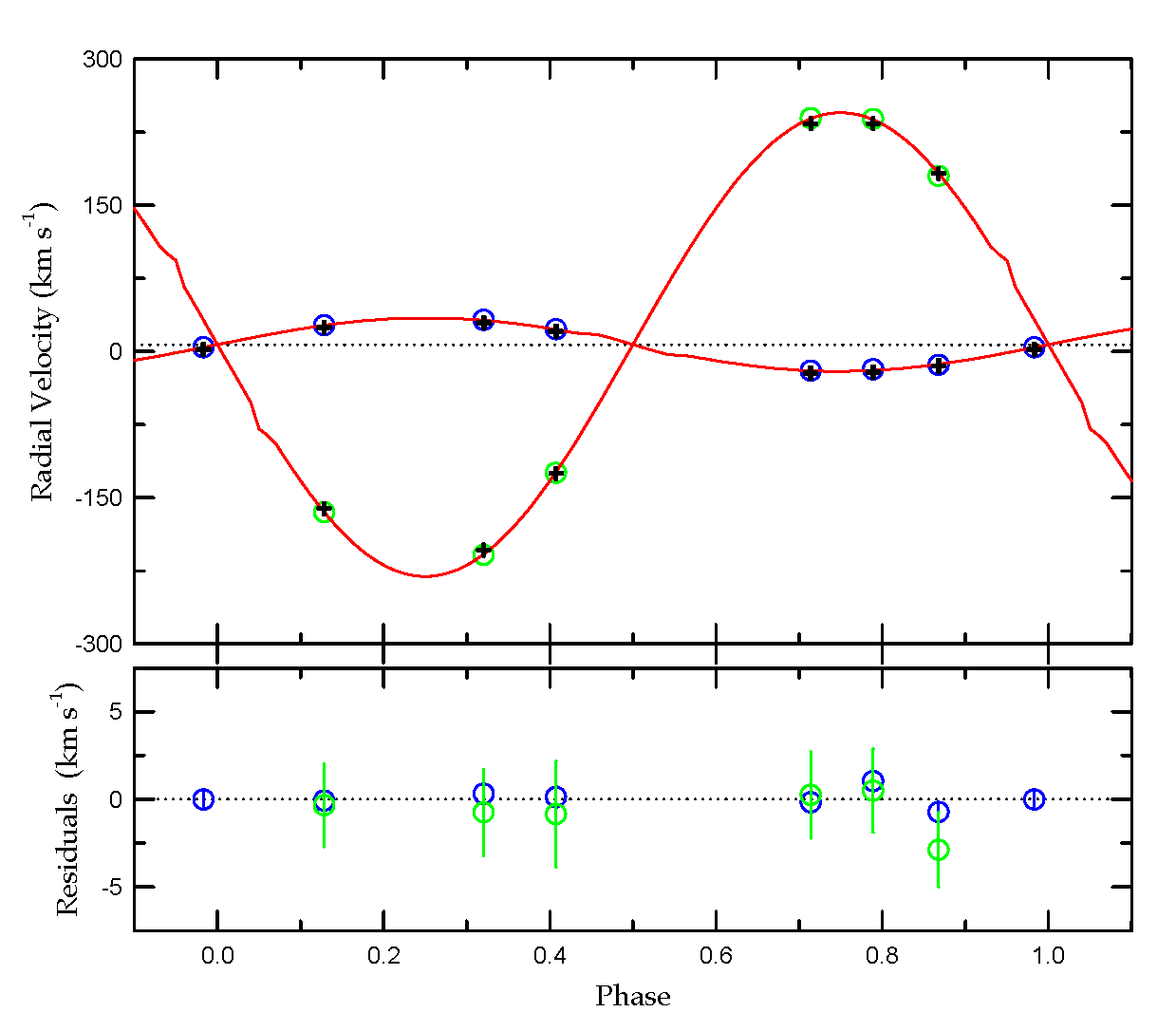}
\caption{RV curves of WASP 1021-28 with fitted models. The blue and green circles are our measurements for the primary (A) and 
secondary (B) stars, respectively. In the upper panel, the solid curves represent the results of the W-D binary model, and 
the dotted line denotes the system velocity of $+$6.98 km s$^{-1}$. The lower panel shows the residuals between our measurements 
and models, where each vertical line is an error bar for each RV. For comparison, the RV measures by \c Cakirli et al. (2024) are 
displayed as black pluses in the upper panel. } 
\label{Fig3}
\end{figure}

\begin{figure}
\includegraphics[]{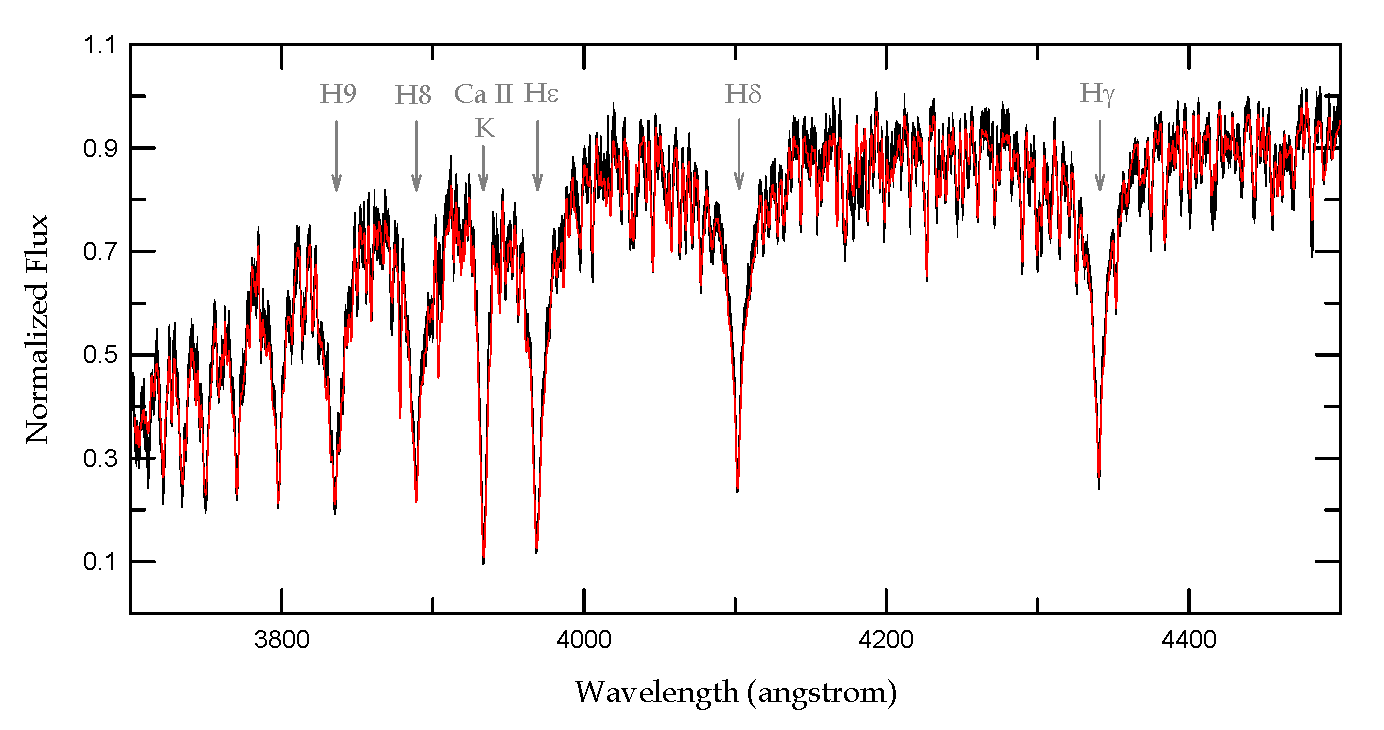}
\caption{Spectrum of WASP 1021-28 observed during the primary eclipse ($\phi$ = 0.983). The black and red solid lines represent 
the observed UVES and best-fitting synthetic spectra, respectively. }
\label{Fig4}
\end{figure}

\begin{figure}
\includegraphics[]{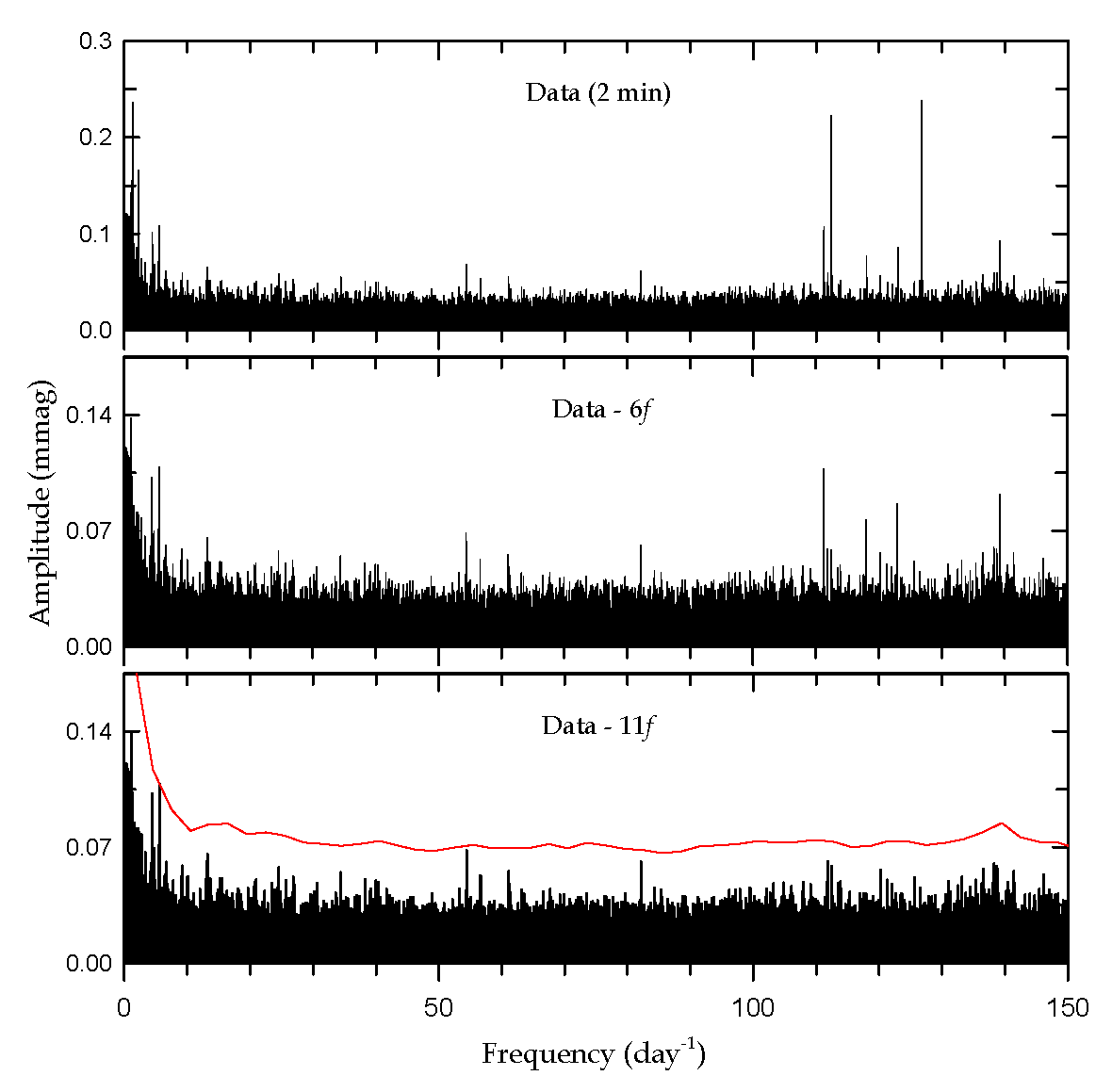}
\caption{Amplitude spectra before (top panel) and after prewhitening the first six frequencies (middle) and all eleven frequencies 
(bottom) from the PERIOD04 program for the entire full-phase 2-min cadence residuals. The red line in the bottom panel corresponds 
to five times the noise spectrum. } 
\label{Fig5}
\end{figure}

\begin{figure}
\includegraphics[]{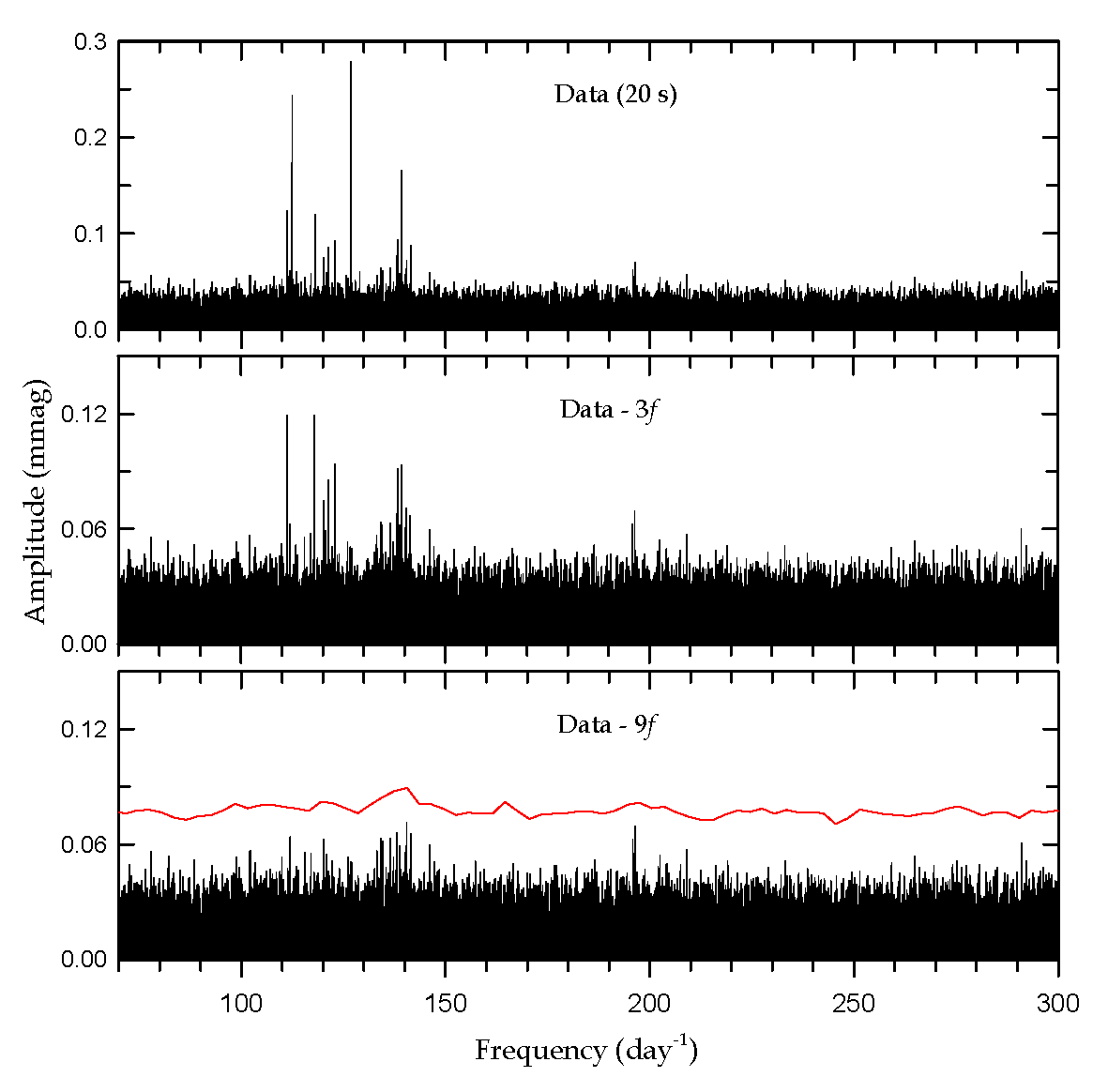}
\caption{PERIOD04 periodograms for all 20-s residual lights, excluding the primary eclipse times. Nine frequencies were found 
in the frequency region between 100 and 150 day$^{-1}$ through iterative prewhitening. The red line in the bottom panel corresponds 
to five times the noise spectrum. } 
\label{Fig6}
\end{figure}

\begin{figure}
\includegraphics[]{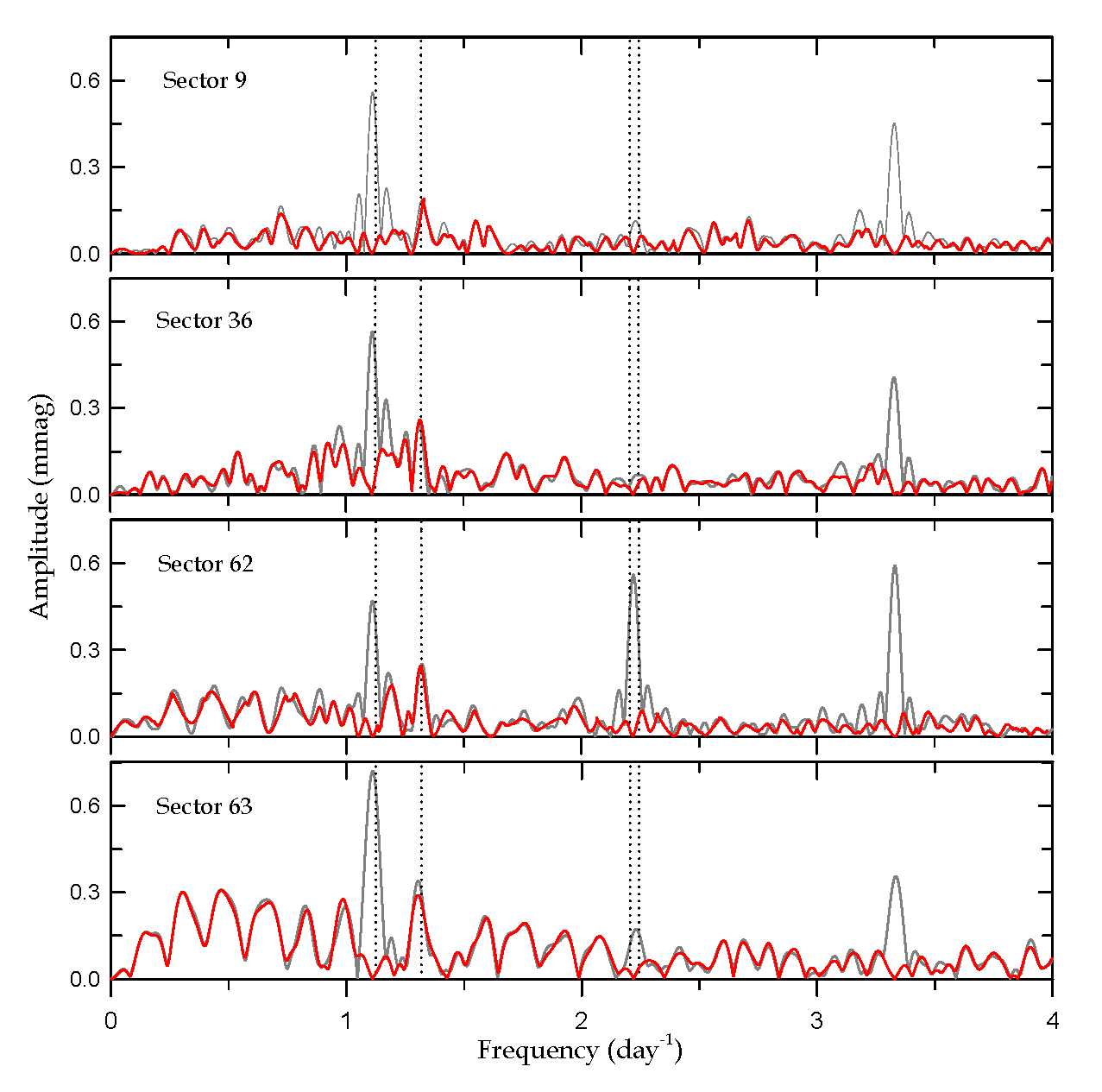}
\caption{Amplitude spectra for the entire TESS data in each sector (S9, S36, S62, and S63). In each panel, the gray and red lines 
are the periodograms for the W-D model residuals and the immune data with orbital frequencies subtracted from them, respectively. 
The vertical dotted lines denote the four low frequencies extracted from the multi-sector data, listed in Table 4. }
\label{Fig7}
\end{figure}

\begin{figure}
\includegraphics[]{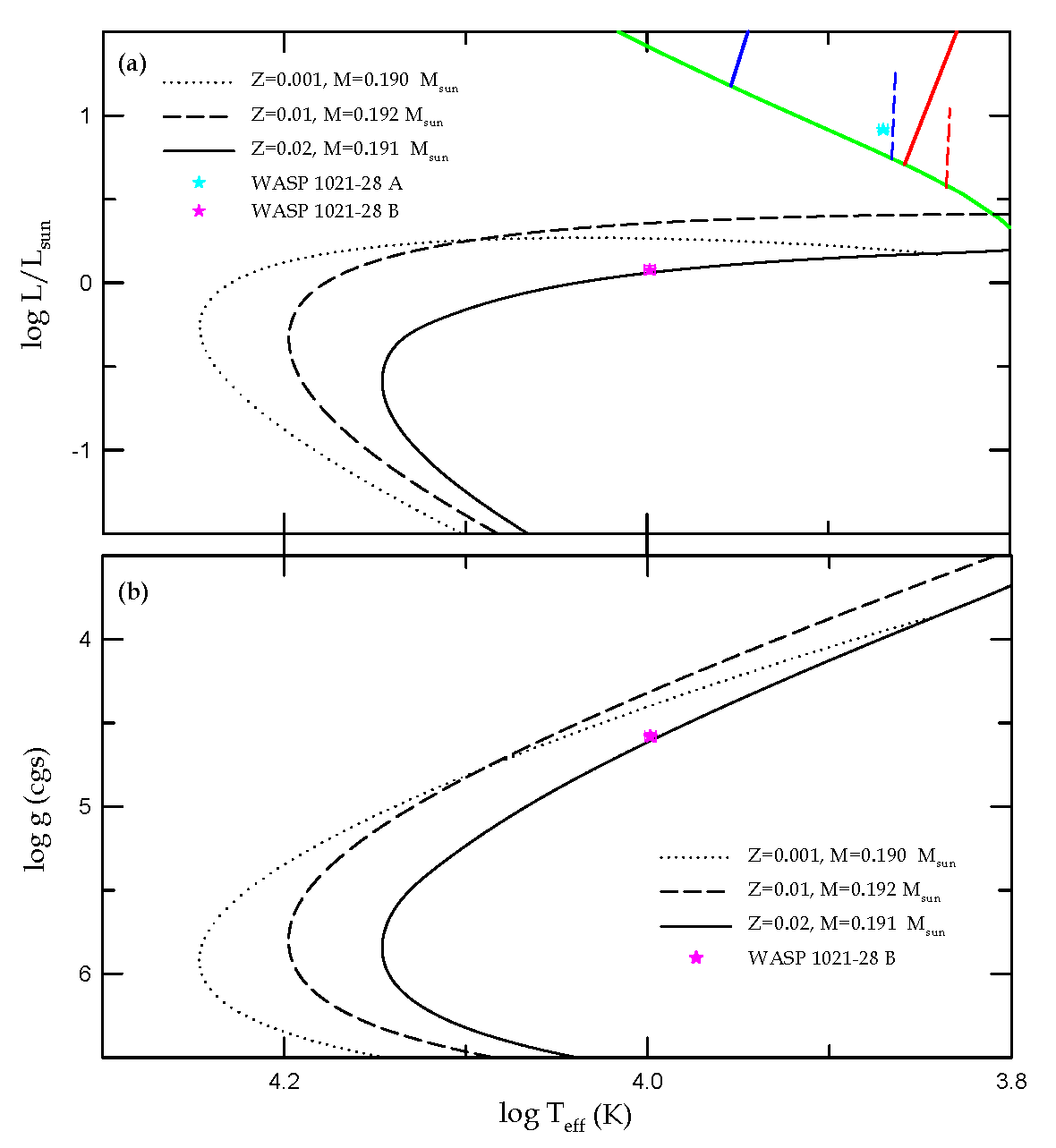}
\caption{(a) H-R and (b) $\log T_{\rm eff}-\log g$ diagrams for WASP 1021-28 (star symbols). In panel (a), the green solid line is 
the zero-age MS, and the colored oblique solid and dash lines denote the instability strips of $\delta$ Sct and $\gamma$ Dor stars, 
respectively (see Lee et al. 2024). In both panels, the black dotted, dashed, and solid lines are the evolutionary tracks of 
low-mass He WDs with metallicities $Z$ of 0.001, 0.01, and 0.02, respectively, for masses of about 0.19 $M_\odot$ (Istrate et al. 2016). } 
\label{Fig8}
\end{figure}

\clearpage
\begin{table}
\caption{TESS Eclipse Timings of WASP 1021-28. A sample is shown here: the full version is provided as supplementary material to the online article. }
\begin{tabular}{lcccc}
\hline
BJD              & Error           & Epoch           & $O-C$           & Min            \\                    
\hline                                         
2,458,544.64284  & $\pm$0.00015    & $-0.5$          & $-$0.00002      & II             \\
2,458,545.09330  & $\pm$0.00008    & $ 0.0$          & $-$0.00001      & I              \\
2,458,545.54347  & $\pm$0.00016    & $ 0.5$          & $-$0.00029      & II             \\
2,458,545.99450  & $\pm$0.00009    & $ 1.0$          & $+$0.00029      & I              \\
2,458,546.44472  & $\pm$0.00027    & $ 1.5$          & $+$0.00006      & II             \\
2,458,546.89497  & $\pm$0.00010    & $ 2.0$          & $-$0.00014      & I              \\
2,458,547.34582  & $\pm$0.00018    & $ 2.5$          & $+$0.00026      & II             \\
2,458,547.79634  & $\pm$0.00010    & $ 3.0$          & $+$0.00034      & I              \\
2,458,548.24629  & $\pm$0.00019    & $ 3.5$          & $-$0.00016      & II             \\
2,458,548.69701  & $\pm$0.00010    & $ 4.0$          & $+$0.00011      & I              \\
\hline
\end{tabular}
\end{table}

\begin{table}
\caption{Radial Velocities of WASP 1021-28$^\dagger$. }
\begin{tabular}{lcrcrc}
\hline
BJD          & Phase   & $V_{\rm A}$   & $\sigma_{\rm A}$ & $V_{\rm B}$   & $\sigma_{\rm B}$ \\ 
(2,450,000+) &         & (km s$^{-1}$) & (km s$^{-1}$)    & (km s$^{-1}$) & (km s$^{-1}$)    \\
\hline 
6983.82076   & 0.9831  & $  4.05$      & 0.54             & $ \dots $     &\dots             \\
6999.79441   & 0.7139  & $-19.96$      & 0.61             & $ 239.01$     & 2.48             \\
7018.85191   & 0.8678  & $-13.92$      & 0.50             & $ 179.81$     & 2.13             \\
7040.70821   & 0.1284  & $ 26.60$      & 0.59             & $-165.01$     & 2.38             \\
7041.86050   & 0.4074  & $ 22.53$      & 0.62             & $-124.70$     & 3.02             \\
7056.61863   & 0.7890  & $-18.48$      & 0.62             & $ 238.28$     & 2.39             \\
7070.61081   & 0.3204  & $ 32.25$      & 0.58             & $-208.86$     & 2.46             \\
\hline
\multicolumn{6}{l}{$^\dagger$ $V_{\rm A}$ and $V_{\rm B}$ are the RV measurements of the primary and secondary stars,} \\
\multicolumn{6}{l}{respectively, and $\sigma_{\rm A}$ and $\sigma_{\rm B}$ are their uncertainties. } \\
\end{tabular}
\end{table}

\begin{table}
\caption{Binary Parameters for WASP 1021-28. }
\begin{tabular}{lccccc}
\hline
Parameter                     & \multicolumn{2}{c}{\c Cakirli et al. (2024)}     && \multicolumn{2}{c}{This Paper}                    \\ [1.0mm] \cline{2-3} \cline{5-6} \\[-2.0ex]
                              & Primary (A)           & Secondary (B)            && Primary (A)           & Secondary (B)             \\
\hline
$T_0$ (BJD)                   & \multicolumn{2}{c}{2,458,544.6430(3)}            && \multicolumn{2}{c}{2,458,545.093506(75)}          \\
$P_{\rm orb}$ (day)           & \multicolumn{2}{c}{0.9008982(3)}                 && \multicolumn{2}{c}{0.900898230(62)}               \\
$i$ (deg)                     & \multicolumn{2}{c}{85.19(4)}                     && \multicolumn{2}{c}{83.216(85)}                    \\
$T_{\rm eff}$ (K)             & 7600(50)              & 10,251(200)              && 7411(40)              & 9965(70)                  \\
$\Omega$                      & 2.995(1)              & 3.220(2)                 && 2.886(21)             & 2.958(31)                 \\
$\Omega_{\rm in}$$\rm ^a$     & \multicolumn{2}{c}{2.007}                        && \multicolumn{2}{c}{2.010}                         \\
$F$                           & 1.0                   & 1.0                      && 0.890(42)             & 1.0                       \\
$x$, $y$                      & 0.337$\rm ^b$         & 0.292$\rm ^b$            && 0.522, 0.261          & 0.410, 0.199              \\
$l/(l_{\rm A}+l_{\rm B}+l_3)$ & 0.806(1)              & 0.074                    && 0.8980(22)            & 0.0729                    \\
$l_3$$\rm ^c$                 & \multicolumn{2}{c}{0.120(1)}                     && \multicolumn{2}{c}{0.0291(20)}                    \\
$r$ (pole)                    & \dots                 & \dots                    && 0.3602(27)            & 0.0770(18)                \\
$r$ (point)                   & \dots                 & \dots                    && 0.3729(31)            & 0.0776(19)                \\
$r$ (side)                    & \dots                 & \dots                    && 0.3681(30)            & 0.0771(18)                \\
$r$ (back)                    & \dots                 & \dots                    && 0.3708(31)            & 0.0776(19)                \\
$r$ (volume)$\rm ^d$          & 0.3562(1)             & 0.07830(7)               && 0.3665(30)            & 0.0772(19)                \\ [1.0mm]   
\multicolumn{6}{l}{Spectroscopic orbits:}                                                                                             \\
$T_0$ (BJD)                   & \multicolumn{2}{c}{\dots}                        && \multicolumn{2}{c}{2,456,983.8361(22)}            \\
$a$ (R$_\odot$)               & \multicolumn{2}{c}{4.646(13)}                    && \multicolumn{2}{c}{4.764(15)}                     \\
$\gamma$ (km s$^{-1}$)        & \multicolumn{2}{c}{5(2)}                         && \multicolumn{2}{c}{6.98(23)}                      \\
$K$ (km s$^{-1}$)             & 27(3)                 & 233(7)                   && 27.82(32)             & 237.95(76)                \\
$q$                           & \multicolumn{2}{c}{0.116(15)}                    && \multicolumn{2}{c}{0.1169(14)}                    \\ [1.0mm]
\multicolumn{6}{l}{Absolute dimensions:}                                                                                              \\
$M$ ($M_\odot$)               & 1.48(13)              & 0.17(3)                  && 1.602(11)             & 0.187(2)                  \\
$R$ ($R_\odot$)               & 1.65(5)               & 0.36(1)                  && 1.746(15)             & 0.368(9)                  \\
$\log$ $g$ (cgs)              & 4.17(1)               & 4.55(5)                  && 4.159(8)              & 4.580(22)                 \\
$\rho$ ($\rho_\odot$)         & \dots                 & \dots                    && 0.302(8)              & 3.78(29)                  \\
$v_{\rm sync}$ (km s$^{-1}$)  & 93(3)                 & 20.5(6)                  && 98.02(86)             & 20.65(51)                 \\
$v$$\sin$$i$ (km s$^{-1}$)    & 82(2)                 & 26(4)                    && 86.6(4.0)             & \dots                     \\
$\log$ $L$ ($L_\odot$)        & 0.91(5)               & 0.12(4)                  && 0.915(12)             & 0.077(25)                 \\
$M_{\rm bol}$ (mag)           & 2.46(13)              & 4.45(10)                 && 2.442(30)             & 4.538(62)                 \\
BC (mag)                      & \dots                 & \dots                    && 0.035(1)              & $-$0.242(15)              \\
$M_{\rm V}$ (mag)             & \dots                 & \dots                    && 2.407(30)             & 4.780(64)                 \\
Distance (pc)                 & \multicolumn{2}{c}{513(19)}                      && \multicolumn{2}{c}{558(9)}                        \\
\hline
\multicolumn{6}{l}{$^a$ Potential for the inner critical Roche surface.} \\
\multicolumn{6}{l}{$^b$ Linear limb-darkening law was used.} \\
\multicolumn{6}{l}{$^c$ Value at 0.25 orbital phase.} \\
\multicolumn{6}{l}{$^d$ Mean volume radius.} \\
\end{tabular}
\end{table}

\clearpage
\begin{table}
\caption{Multi-frequency Solution for WASP 1021-28$\rm ^{a,b}$. }
\begin{tabular}{lrccrc}
\hline
             & Frequency              & Amplitude           & Phase           & SNR$\rm ^c$    & Remark          \\
             & (day$^{-1}$)           & (mmag)              & (rad)           &                &                 \\
\hline
$f_{1}$      &   1.12495$\pm$0.00004  & 0.188$\pm$0.056     & 1.34$\pm$0.87   &  5.77          &                 \\   
$f_{2}$      &   1.31865$\pm$0.00003  & 0.237$\pm$0.060     & 5.61$\pm$0.74   &  6.77          & $\gamma$ Dor    \\   
$f_{3}$      &   2.20517$\pm$0.00003  & 0.204$\pm$0.051     & 6.16$\pm$0.74   &  6.79          &                 \\   
$f_{4}$      &   2.24333$\pm$0.00004  & 0.179$\pm$0.051     & 5.00$\pm$0.83   &  6.03          &                 \\   
$f_{5}$      & 111.25132$\pm$0.00003  & 0.119$\pm$0.027     & 2.66$\pm$0.67   &  7.52          & pre-ELMV        \\   
$f_{6}$      & 112.42776$\pm$0.00002  & 0.241$\pm$0.027     & 1.33$\pm$0.33   & 15.16          & pre-ELMV        \\   
$f_{7}$      & 118.01016$\pm$0.00003  & 0.116$\pm$0.027     & 3.64$\pm$0.68   &  7.35          & pre-ELMV        \\   
$f_{8}$      & 121.33540$\pm$0.00005  & 0.082$\pm$0.028     & 1.99$\pm$1.01   &  4.96          & pre-ELMV        \\   
$f_{9}$      & 122.98200$\pm$0.00004  & 0.094$\pm$0.027     & 2.70$\pm$0.85   &  5.87          & pre-ELMV        \\   
$f_{10}$     & 126.74621$\pm$0.00001  & 0.280$\pm$0.026     & 5.74$\pm$0.27   & 18.31          & pre-ELMV        \\   
$f_{11}$     & 138.30988$\pm$0.00005  & 0.092$\pm$0.031     & 1.58$\pm$1.00   &  5.04          & pre-ELMV        \\   
$f_{12}$     & 139.22954$\pm$0.00004  & 0.103$\pm$0.031     & 1.02$\pm$0.89   &  5.64          & pre-ELMV        \\   
$f_{13}$     & 139.23163$\pm$0.00002  & 0.194$\pm$0.031     & 5.61$\pm$0.47   & 10.69          & pre-ELMV        \\   
\hline 
\multicolumn{6}{l}{$^a$ Listed in order of frequency.} \\
\multicolumn{6}{l}{$^b$ Parameters' errors were obtained following Kallinger, Reegen \& Weiss (2008).} \\
\multicolumn{6}{l}{$^c$ Calculated in a range of 5 day$^{-1}$ around each frequency.} \\
\end{tabular}
\end{table}

\bsp
\label{lastpage}

\begin{thebibliography}{99}
\bibitem[\protect\citeauthoryear{Baran \& Koen}{2021}]{b1} Baran A. S., Koen C., 2021, AcA, 71, 113 
\bibitem[\protect\citeauthoryear{Beuzit et al}{2008}]{b2} Beuzit J.-L. et al., 2008, Proc. SPIE, 7014, 701418
\bibitem[\protect\citeauthoryear{Blanco-Cuaresma et al}{2014}]{b3} Blanco-Cuaresma S. et al., 2014, A\&A, 569, A111
\bibitem[\protect\citeauthoryear{Bowman \& Michielsen}{2021}]{b4} Bowman D. M., Michielsen M., 2021, A\&A, 656, A15
\bibitem[\protect\citeauthoryear{Cakirli et al}{2024}]{b5} \c Cakirli \" O., Hoyman B., O\"zdarcan O., 2024, MNRAS, 533, 2058
\bibitem[\protect\citeauthoryear{Coelho et al}{2005}]{b6} Coelho P., Barbuy B., Melendez J., Sciavon R. P., Castilho B. V., 2005, A\&A, 443, 735
\bibitem[\protect\citeauthoryear{Dekker et al}{2000}]{b7} Dekker H. et al., 2000, Proc. SPIE, 4008, 534 
\bibitem[\protect\citeauthoryear{Dohlen et al}{2008}]{b8} Dohlen K. et al., 2008, Proc. SPIE, 7014, 70143L
\bibitem[\protect\citeauthoryear{Flower}{1996}]{b9} Flower P. J., 1996, ApJ, 469, 355
\bibitem[\protect\citeauthoryear{Fuller \& Felce}{2023}]{b10} Fuller J., Felce C., 2023, MNRAS, 527, L103
\bibitem[\protect\citeauthoryear{GAIA}{2022}]{b11} Gaia Collaboration 2022, VizieR Online Data Catalog, I/355
\bibitem[\protect\citeauthoryear{Guerri et al}{2011}]{b12} Guerri G. et al., 2011, ExA, 30, 59
\bibitem[\protect\citeauthoryear{Guo}{2021}]{b13} Guo Z., 2021, FrASS, 8, 67
\bibitem[\protect\citeauthoryear{Hong et al}{2021}]{b14} Hong K. et al., 2021, AJ, 161, 137
\bibitem[\protect\citeauthoryear{Istrate et al}{2016}]{b15} Istrate A. G. et al., 2016, A\&A, 595, A35
\bibitem[\protect\citeauthoryear{Kallinger et al}{2008}]{b16} Kallinger T., Reegen P., Weiss W. W., 2008, A\&A, 481, 571
\bibitem[\protect\citeauthoryear{Kallrath}{2022}]{b17} Kallrath J., 2022, Galaxies, 10, 17 
\bibitem[\protect\citeauthoryear{Kwee \& van Woerden}{1956}]{b18} Kwee K. K., van Woerden H., 1956, BAN, 12, 327
\bibitem[\protect\citeauthoryear{Lagos et al}{2020}]{b19} Lagos F. et al., 2020, MNRAS, 499, L121
\bibitem[\protect\citeauthoryear{Lee et al}{2024}]{b20} Lee J. W., Hong K., Jeong M.-J., Wolf M., 2024, ApJ, 973, 113
\bibitem[\protect\citeauthoryear{Lee et al}{2017}]{b21} Lee J. W., Hong K., Kim S.-L., Koo J.-R., 2017, ApJ, 835, 189
\bibitem[\protect\citeauthoryear{Lee et al}{2023}]{b22} Lee J. W., Hong K., Park J.-H., Wolf M., Kim D.-J., 2023, AJ, 165, 159
\bibitem[\protect\citeauthoryear{Lee et al}{2014}]{b23} Lee J. W., Kim S.-L., Hong K., Lee C.-U., Koo J.-R., 2014, AJ, 148, 37
\bibitem[\protect\citeauthoryear{Lee et al}{2020}]{b24} Lee J. W., Koo J.-R., Hong K., Park J.-H., 2020, AJ, 160, 49 
\bibitem[\protect\citeauthoryear{Lenz \& Breger}{2005}]{b25} Lenz P., Breger M., 2005, Comm. Asteroseismology, 146, 53
\bibitem[\protect\citeauthoryear{Maxted et al}{2014a}]{b26} Maxted P. F. L. et al., 2014a, MNRAS, 437, 1681
\bibitem[\protect\citeauthoryear{Maxted et al}{2014b}]{b27} Maxted P. F. L. et al., 2014b, MNRAS, 444, 208
\bibitem[\protect\citeauthoryear{Maxted et al}{2013}]{b28} Maxted P. F. L. et al., 2013, Natur, 498, 463
\bibitem[\protect\citeauthoryear{Paegert}{2022}]{b30} Paegert M. et al., 2022, VizieR Online Data Catalog, IV/39 
\bibitem[\protect\citeauthoryear{Pauli et al}{2006}]{b31} Pauli E.-M., Napiwotzki R., Heber U., Altmann M., Odenkirchen M., 2006, A\&A, 447, 173
\bibitem[\protect\citeauthoryear{Pecaut \& Mamajek}{2013}]{b32} Pecaut M. J., Mamajek E. E. 2013, ApJS, 208, 9
\bibitem[\protect\citeauthoryear{Pilecki et al}{2017}]{b33} Pilecki B. et al., 2017, ApJ, 842, 110
\bibitem[\protect\citeauthoryear{Ricker et al}{2015}]{b34} Ricker G. R. et al., 2015, JATIS, 1, 014003
\bibitem[\protect\citeauthoryear{Rucinski}{2002}]{b36} Rucinski S. M., 2002, AJ, 124, 1746 
\bibitem[\protect\citeauthoryear{Rucinski}{2004}]{b37} Rucinski S. M., 2004, in IAU Symp. 215, Stellar Rotation, ed. A. Maeder \& P. Eenens (San Francisco, CA: ASP), 17
\bibitem[\protect\citeauthoryear{Southworth \& Bowman}{2022}]{b38} Southworth J., Bowman D. M., 2022, The Observatory, 142, 161
\bibitem[\protect\citeauthoryear{Tkachenko}{2015}]{b40} Tkachenko A., 2015, A\&A, 581, A129
\bibitem[\protect\citeauthoryear{Torres}{2010}]{b41} Torres G., 2010, AJ, 140, 1158
\bibitem[\protect\citeauthoryear{van Hamme}{1993}]{b42} van Hamme W., 1993, AJ, 106, 209
\bibitem[\protect\citeauthoryear{van Hamme \& Wilson}{2007}]{b43} van Hamme W., Wilson R. E., 2007, ApJ, 661, 1129
\bibitem[\protect\citeauthoryear{van Roestel et al}{2018}]{b44} van Roestel J. et al., 2018, MNRAS, 475, 2560
\bibitem[\protect\citeauthoryear{Vigan et al}{2010}]{b45} Vigan A. et al., 2010, in In the Spirit of Lyot 2010. p. E48
\bibitem[\protect\citeauthoryear{Wang et al}{2020}]{b46} Wang K., Zhang X., Dai M., 2020, ApJ, 888, 49
\bibitem[\protect\citeauthoryear{Wilson \& Devinney}{1971}]{b47} Wilson R. E., Devinney E. J., 1971, ApJ, 166, 605
\end{thebibliography}
\end{document}